\documentclass[
preprintnumbers,
nofootinbib,
amsmath,amssymb,
prd,aps,superscriptaddress,twocolumn,floatfix,
]{revtex4-2}
\usepackage{graphicx}
\usepackage{amsmath}
\usepackage[caption=false]{subfig}
\usepackage{siunitx}
\usepackage{placeins}
\usepackage{color}
\usepackage{standalone}
\usepackage{dcolumn}
\usepackage{tensor}
\usepackage{bm}
\usepackage{microtype}
\usepackage{etoolbox}
\usepackage{amssymb}
\usepackage{mathrsfs}
\usepackage{accents}
\usepackage[normalem]{ulem}
\usepackage{soul} 
\usepackage[dvipsnames]{xcolor}
\usepackage[colorlinks,urlcolor=NavyBlue,citecolor=NavyBlue,linkcolor=NavyBlue,pdfusetitle]{hyperref}
\usepackage[all]{hypcap}
\usepackage[inline]{enumitem}
\usepackage[utf8]{inputenc}
\usepackage{xspace}
\usepackage[printonlyused, nolist]{acronym}
\usepackage{lineno}

\newcommand{\Msun}{\ensuremath{M_{\odot}}\xspace}

\newcommand{\chieff}{\ensuremath{\chi_{\rm eff}}\xspace}
\newcommand{\Xcl}{\ensuremath{\Lambda^{\rm cl}}\xspace}
\newcommand{\xcl}{\ensuremath{\lambda^{\rm cl}}\xspace}
\newcommand{\XclChosen}{\ensuremath{\Tilde{\Lambda}^{\rm cl}}\xspace}
\newcommand{\xbbh}{\ensuremath{\theta^{\rm B}}\xspace}
\newcommand{\Nbbh}{\ensuremath{N^{\rm B}}\xspace}
\newcommand{\etabbh}{\ensuremath{\eta}\xspace}
\newcommand{\Ncl}{\ensuremath{N^{\rm cl}}\xspace}
\newcommand{\Mcl}{\ensuremath{M_{\rm cl}}\xspace}
\newcommand{\rh}{\ensuremath{r_{h}}\xspace}
\newcommand{\aEj}{\ensuremath{a_{\rm ej}}\xspace}

\newcommand{\jhu}{William H. Miller III Department of Physics and Astronomy, Johns Hopkins University, Baltimore, Maryland 21218, USA}

\graphicspath{{./figures/}}

\begin{document}
\title{Single-event likelihood of star cluster properties with LIGO-Virgo-Kagra binary black hole observations}

\author{Ken K.~Y.~Ng}
\email{kng15@jhu.edu}
\author{Konstantinos Kritos}
\author{Andrea Antonelli}
\author{Roberto Cotesta}
\author{Emanuele Berti}

\affiliation{\jhu}

\date{\today}

\begin{abstract}
The population of binary black hole mergers observed in gravitational waves, together with astrophysical simulations, can help us to understand the properties of the progenitors and the binary formation mechanisms in different astrophysical scenarios.
Here we focus on dynamical formation in star clusters. We use the third gravitational-wave transient catalog (GWTC-3) and \textsc{Rapster}, a rapid code to simulate cluster dynamics, to show that it is possible to construct the single-event likelihood of star cluster properties from individual observations. We find that the measured primary mass in a binary black hole merger correlates with the measured star cluster mass, because the mass spectrum of the primary component increases with the mass of the cluster.
This trend may be caused by two physical mechanisms: (i) the more efficient production of hierarchical mergers with primary mass above $\sim 40~\Msun$ for cluster masses of $\gtrsim 10^6~\Msun$, and (ii) the suppression of more massive first-generation binaries, which happens because ejected binaries do not merge within the lookback time for cluster masses of $\lesssim 10^5~\Msun$.
The formalism presented here can be generalized to infer the population properties of binary progenitors in more realistic scenarios involving multiple formation channels.
\end{abstract}

\maketitle
\section{Introduction}

The recent catalogs of gravitational-wave (GW) transients released by the LIGO-Virgo-Kagra Collaboration~\cite{GWTC1,GWTC2,GWTC3,aLIGO,aVirgo,kagra} motivated efforts to investigate the properties of the binary black hole (BBH)  population and their possible formation channels. There are various ways to address this problem.

One approach is to construct {\em phenomenological models} that reproduce the main distinctive features of astrophysical formation channels~\cite{GWTC1rate,GWTC2rate,GWTC3rate}.
This is a sensible approach because it requires minimal astrophysical modeling.

For instance, certain broad features of the population -- such as the BH spin alignment with the orbital angular momentum~\cite{Gerosa:2013laa,Vitale:2015tea,Farr:2017gtv,Farr:2017uvj,Ng:2018neg,Talbot:2017yur,Wysocki:2017isg,Fishbach:2022lzq,Fishbach:2023xws} or the measurement of binary component masses populating the pair instability supernova (PISN) mass gap, above $\sim 40\Msun$~\cite{Gerosa:2017kvu,Fishbach:2017dwv,Talbot:2018cva,Farah:2023vsc,Kimball:2020opk} -- may provide evidence for multiple formation channels. This is because isolated binary evolution is expected to produce mostly binaries with aligned spins and masses below the mass gap~\cite{Belczynski:2005mr,Belczynski:2016obo,Belczynski:2017gds,Dominik:2012kk,Dominik:2013tma,Dominik:2014yma,Bavera:2019fkg,Breivik:2019lmt,Stevenson:2017tfq,Broekgaarden:2019qnw,Baibhav:2019gxm,deMink:2015yea,Belczynski:2016jno,Woosley:2016hmi,COMPASTeam:2021tbl}, while the spins of BHs produced through dynamical formation in star clusters should be isotropically oriented, and hierarchical mergers can populate the PISN mass gap~\cite{Askar:2016jwt,Rodriguez:2021qhl,Mapelli:2021gyv,Antonini:2019ulv,Sedda:2021vjh,Kritos:2022ggc,Rodriguez:2019huv,Rodriguez:2015oxa,Rodriguez:2016kxx,Rodriguez:2017pec,Kritos:2022ggc,Doctor:2019ruh,DiCarlo:2019pmf,Antonini:2020xnd,Kremer:2020wtp,Gerosa:2021mno}. Other features that can be captured by phenomenological models include the time (or redshift) evolution of the merger rate density~\cite{Fishbach:2018edt,Fishbach:2021mhp,Mapelli:2019bnp,Rodriguez:2018rmd,Ng:2020qpk,Vitale:2018yhm}, or the presence of peaks and tails in the redshift evolution of merger rate densities due to putative Population~III or primordial BBH components, which could be observable with next-generation GW detectors~\cite{Ng:2020qpk,Ng:2022agi}.

One drawback of phenomenological models is that they are affected by modeling systematics: for example, the class of parametrized functions used to reproduce the data may be too restrictive, leading to an erroneous mapping between the parametrized models and detailed astrophysical simulations.

A second approach is to infer the empirical distribution using {\em data-driven models}, leaving the interpretation of the resulting distribution to the postprocessing stage~\cite{Mandel:2016prl,Rinaldi:2021bhm,Edelman:2022ydv,Wong:2022bxp}. Even within this approach, finding a suitable statistical metric connecting data with astrophysical simulations could be problematic.

A third approach (and one that we follow in this paper) is to build a {\em direct mapping} between the measured parameters of a BBH merger event
and the observables predicted by astrophysical simulations~\cite{Zevin:2017evb,Wong:2020jdt,Wong:2022flg,Andrews:2017ads,Taylor:2018iat,Mould:2022ccw,Mould:2023ift,Antonelli:2023gpu}.
While the inference is still limited by our incomplete knowledge of astrophysical formation scenarios, this approach allows for in-depth studies of the astrophysical mechanisms that correspond to certain features seen in the populations.
There have been many attempts to infer
some of the key astrophysical parameters affecting the isolated binary evolution scenario, as well as the relative contribution (or branching ratios) of multiple formation channels: see e.g.~\cite{Stevenson:2015bqa,Bouffanais:2019nrw,Wong:2020ise,Wong:2020yig,Zevin:2020gbd,Andrews:2020pjg} for an incomplete list.

In this paper we avoid the complications related to multiple formation channels, and we focus on the dynamical formation scenario in dense star clusters.
Each cluster in the star cluster population has different properties, and therefore it produces a different BBH subpopulation. Here we develop a two-level hierarchical Bayesian framework that can ultimately infer the properties of the star cluster population from BBH merger observations (Sec.~\ref{sec:Bayes}).
We focus on the first step in this framework, which consists in constructing the single-event likelihood of star cluster properties: i.e., we aim to identify clusters with parameters which are more likely to generate a particular BBH observed in the third GW transient catalog (GWTC-3)~\cite{GWTC3,gwoscO3}.
To this end, we use a code for rapid simulations of cluster dynamics, \textsc{Rapster}~\cite{Kritos:2022ggc}, to build a statistical mapping between the BBH parameters and the star cluster parameters.
By analyzing these simulations we observe a positive correlation between the measured BBH primary mass and the inferred cluster mass.
As we discuss in Sec.~\ref{sec:clusterlikelihood}, this correlation may be explained by the cluster mass scaling of the efficiency in the production of hierarchical mergers and by the inspiral timescale of the ejected binaries.
In Sec.~\ref{sec:discussion} we discuss some technical aspects and future prospects to interpret the observed BBH population using astrophysical simulations.
In Appendix~\ref{App:KDEandSampling} we give details on the cluster simulations and on the kernel density estimation (KDE) we use to approximate the joint distribution from the simulated mergers.

\section{Two-level hierarchical Bayesian framework}
\label{sec:Bayes}

As the Universe evolves, numerous star clusters form with redshift-dependent rates and with different physical properties (such as total mass, radius, and metallicity)~\cite{2019ARA&A..57..227K}.
Each cluster evolves dynamically and produces an ensemble of BBHs whose statistical distribution depends on the properties of the host cluster.
Therefore, the distribution of BBH properties observed by LVK in the cluster scenario should be modeled by considering the population of BBHs originating from \textit{a population of star clusters}.
The ``inverse problem'' consists of inferring the properties of the star cluster population that can host BBH mergers observable in GW detectors.

One may attempt to perform the full hierarchical analysis by simulating BBHs drawing from different realizations of cluster populations.
However, it is more beneficial to consider a two-level hierarchical model that can break down the inference, as follows.

In the first level of hierarchy, we map the single-event likelihood of BBH parameters to the likelihood of parameters of individual clusters that are likely to produce them, using the BBH properties predicted by star cluster simulations.
In the second level of hierarchy, we combine these single-event cluster likelihoods and infer the distribution of the cluster properties.

To see how the single-event cluster likelihood enters the hierarchical framework, we derive it using a top-down approach.
The full hierarchical likelihood based on a Poisson process of data generation is given by~\cite{Mandel:2018mve,Vitale:2020aaz}
\begin{align}\label{eq:fullHBA}
    p(\Xcl|\pmb{d}) &\propto e^{-\Nbbh_{\rm det}(\Xcl)} \prod_{i=1}^N \int p(d_i | \xbbh_i) \frac{d\Nbbh}{d\xbbh}(\xbbh_i | \Xcl) d\xbbh_i,
\end{align}
where $\pmb{d}=\left\{d_i\right\}_{i=1}^N$ is the data set of $N$ BBH observations, $p(d_i | \xbbh_i)$ is the individual likelihood of the $i$-th BBH characterized by parameters $\xbbh_i$ such as component masses and spins, $d\Nbbh/d\xbbh$ is the differential number of BBHs expected for a given cluster population characterized by hyperparameters $\Xcl$, and $\Nbbh_{\rm det}$ is the number of detectable BBHs:
\begin{align}
    \Nbbh_{\rm det}(\Xcl) = \int \frac{d\Nbbh}{d\xbbh}(\xbbh | \Xcl) \epsilon_{\rm det}(\xbbh) d\xbbh,
\end{align}
where $0\leq \epsilon_{\rm det}(\xbbh)\leq 1$ is the detection efficiency for a BBH merger with binary parameters $\xbbh$.

The differential rate can be written as
\begin{align}
    & \indent
    \frac{d\Nbbh}{d\xbbh}(\xbbh | \Xcl)
    = \int \frac{d^2 \Nbbh}{d\xbbh d\xcl}(\xbbh , \xcl| \Xcl) d\xcl \nonumber\\
    & = \int p(\xbbh | \xcl) \etabbh(\xcl) \frac{d\Ncl}{d\xcl}(\xcl | \Xcl) d\xcl,
\end{align}
where $p(\xbbh | \xcl)$ is the distribution of $\xbbh$ originating from a single cluster characterized by some $\xcl$, $\etabbh(\xcl)$ is the number of BBHs produced by the cluster, and $\Ncl$ is the total number of clusters.
For example, $\xcl$ could be the mass of a single cluster, and $\Xcl$ the power law index of the cluster mass function.
Both $p(\xbbh | \xcl)$ and $\etabbh(\xcl)$ are predicted by the simulation, while $p(d_i | \xbbh_i)$ is obtained from GW observations.

The integral in Eq.~\eqref{eq:fullHBA} is equivalent to the expected number of BBHs averaged over the individual BBH likelihoods. It can be rewritten as
\begin{align}\label{eq:Nexpected}
    \indent &\langle \Nbbh \rangle_i (\Xcl) \nonumber \\
    = & \iint p(d_i | \xbbh_i) p(\xbbh_i | \xcl_i) \etabbh(\xcl_i) \frac{d\Ncl}{d\xcl}(\xcl_i | \Xcl) d\xcl_i d\xbbh_i \nonumber \\
    = & \int p(d_i | \xcl_i) \etabbh(\xcl_i)\frac{d\Ncl}{d\xcl}(\xcl_i | \Xcl) d\xcl_i,
\end{align}
where the individual cluster likelihood of the $i$-th observation, $p(d_i | \xcl_i)$, is the marginalization of $p(\xbbh_i | \xcl_i)$ over the individual BBH likelihoods:
\begin{align}\label{eq:clusterLikelihood}
    p(d_i | \xcl_i) \equiv \int p(d_i | \xbbh_i) p(\xbbh_i | \xcl_i) d\xbbh_i.
\end{align}
This result may also be obtained by applying Bayes' theorem and marginalizing over $\xbbh_i$ on the joint distribution $p(d_i, \xbbh_i | \xcl_i)$ ``directly'' in the bottom-up approach.

This procedure is practically advantageous as we only need to approximate $p(\xbbh | \xcl)$ and $\etabbh(\xcl)$ once with the finite samples produced by the simulations.
On the contrary, the emulation of the entire BBH population $d\Nbbh/d\xbbh$ originating from all possible cluster populations is limited to the choices of the prior functions used in the training set, and thus hinders the use of the more flexible nonparametric models for the cluster population that we are interested in.
In the following, we will study solely the single-event cluster likelihood in Eq.~\eqref{eq:clusterLikelihood} for selected events in GWTC-3.

\section{Individual likelihood of cluster properties}
\label{sec:clusterlikelihood}

The best-measured BBH parameters in current GW observations, and therefore the parameters that are most informative in the inference of formation channels, are the (source-frame) masses of the primary, $m_1$, and secondary, $m_2$; the effective spin projected along the orbital angular momentum, $\chieff$; and the redshift, $z$.

The \textsc{Rapster} code has a total of 19 input parameters. While it is hopeless to constrain all of these parameters, as a proof of principle we explore two of the most important intrinsic properties of individual clusters: the total mass of the cluster, $\Mcl$, and the half-mass radius at the time of cluster formation, $\rh$.
In other words, we set $\xbbh = (m_1, m_2, \chieff, z)$ and $\xcl=(\Mcl, \rh)$ in the formalism of Sec.~\ref{sec:Bayes}.
We reweigh the LVK posterior samples and obtain the likelihood samples of \xbbh.

We limit the cluster parameter space to the ranges $\Mcl \in [10^4,~10^7]~\Msun$ and $\rh \in [0.5,~3]~\rm pc$, respectively, based on current observations of young star clusters~\cite{2019ARA&A..57..227K}.
The simulation samples for constructing the KDE are generated by the following settings.
The initial cluster masses $M_{\rm cl}$ are drawn from a power-law distribution with a spectral index $-2$ in the range $[10^{3.7},\,10^{7.3}] \Msun$. To avoid hard cutoffs in the range $[10^4,~10^7]\Msun$ where we construct our KDE, we taper the distribution using a Tukey window function with shape parameter 0.18.
The initial half-mass radius $r_{h}$ is drawn from a linear distribution in the range $[0.3,\,3]$~pc.
This choice is to balance the number of mergers per cluster in the simulation set, which scales with the inverse of the cluster radius.
In the inference, we obtain the likelihood of \Mcl and \rh by reweighing the chosen priors.
We note that the above initial conditions are reweighed out eventually and that they do not affect the evaluation of the likelihood, as shown in Eq.~\eqref{eq:clusterLikelihood}.

The other cluster parameters and the initial cluster conditions are fixed as follows.
We use \textsc{SEVN} to compute the initial mass function of BHs so that the PISN cut-off is at $\sim 40\Msun$, with the exact value depending on the metallicity~\cite{Spera:2017fyx}.
The dimensionless natal spin of first-generation BHs is sampled from a uniform distribution in the range $[0,~0.2]$, as in Ref.~\cite{Baibhav:2021qzw}.
The masses and spins of BBH merger remnants are computed using the \textsc{precession} code~\cite{Gerosa:2016sys}.
The initial central stellar density is calculated as $3\Mcl/(4\pi(\rh/1.3)^3)$, assuming a Plummer profile~\cite{Plummer:1911zza}.
This choice of mass-radius relation is motivated by observations of star clusters in the local Universe~\cite{2019ARA&A..57..227K}.
A detailed cross-validation of the cluster population resulting from this assumption with other surveys would be interesting, but is beyond the scope of this paper.

Moreover, we assume a young massive cluster population with the redshift of cluster formation and mean metallicity sampled from the Madau-Fragos distribution~\cite{Madau:2016jbv}.
We also apply a log-normal spread with a variance of 0.3 in the metallicity distribution at each redshift.
The rest of the input parameters are set to their default \textsc{Rapster} values, as listed in Table~I of Ref.~\cite{Kritos:2022ggc}.
With one node (48 processors) at the Maryland Advanced Research Computing Center at Johns Hopkins, we can simulate $\sim 10^6$ star clusters within 2 days.

To approximate the conditional probability distribution $p(\xbbh|\xcl)$, we employ Gaussian KDE on a set of $\sim 7\times 10^5$ simulated BBH mergers generated by the synthesis code \textsc{Rapster}.
Given $p(\xbbh|\xcl)$, we can evaluate Eq.~\eqref{eq:clusterLikelihood} for each BBH observation in GWTC-3 released by the LVK Collaboration.
Since the integral is generally intractable, we sample the likelihood in Eq.~\eqref{eq:clusterLikelihood} by Monte Carlo methods.
Technical details on the KDE and on the integration are given in Appendix~\ref{App:KDEandSampling}.

\begin{figure}[t]
    \centering
    \includegraphics[width=0.95\columnwidth]{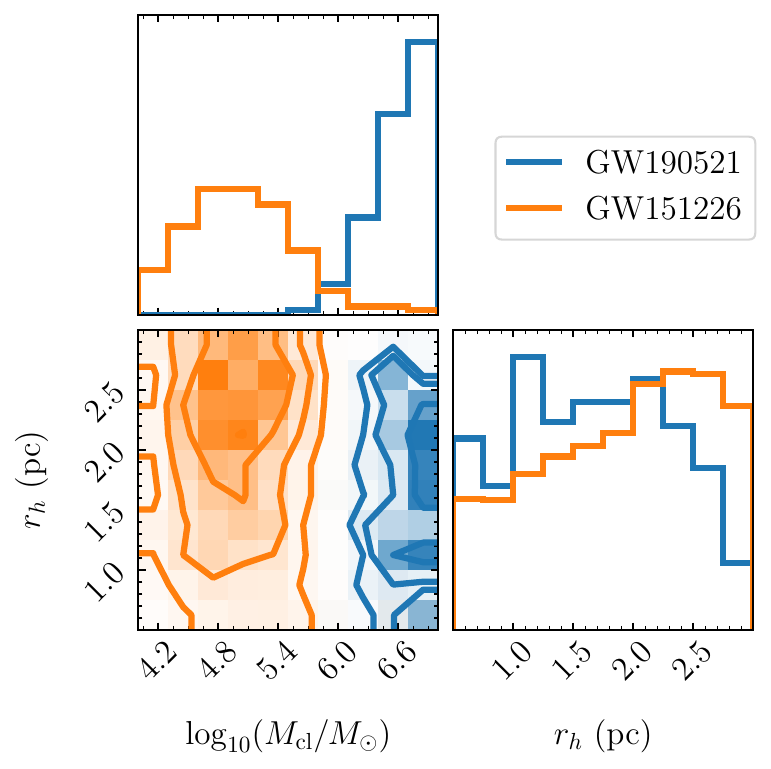}
    \caption{The joint likelihoods of $(\Mcl, \rh)$ for GW190521 (blue) and GW151226 (orange).
    The marginalized $\Mcl$ likelihood for GW190521 (GW151226) favors values above (below) $\Mcl \sim 10^6~\Msun$, while the $\rh$ likelihoods are mostly flat.
    }
    \label{fig:corner}
\end{figure}

\begin{figure}[t]
    \centering
    \includegraphics[width=0.95\columnwidth]{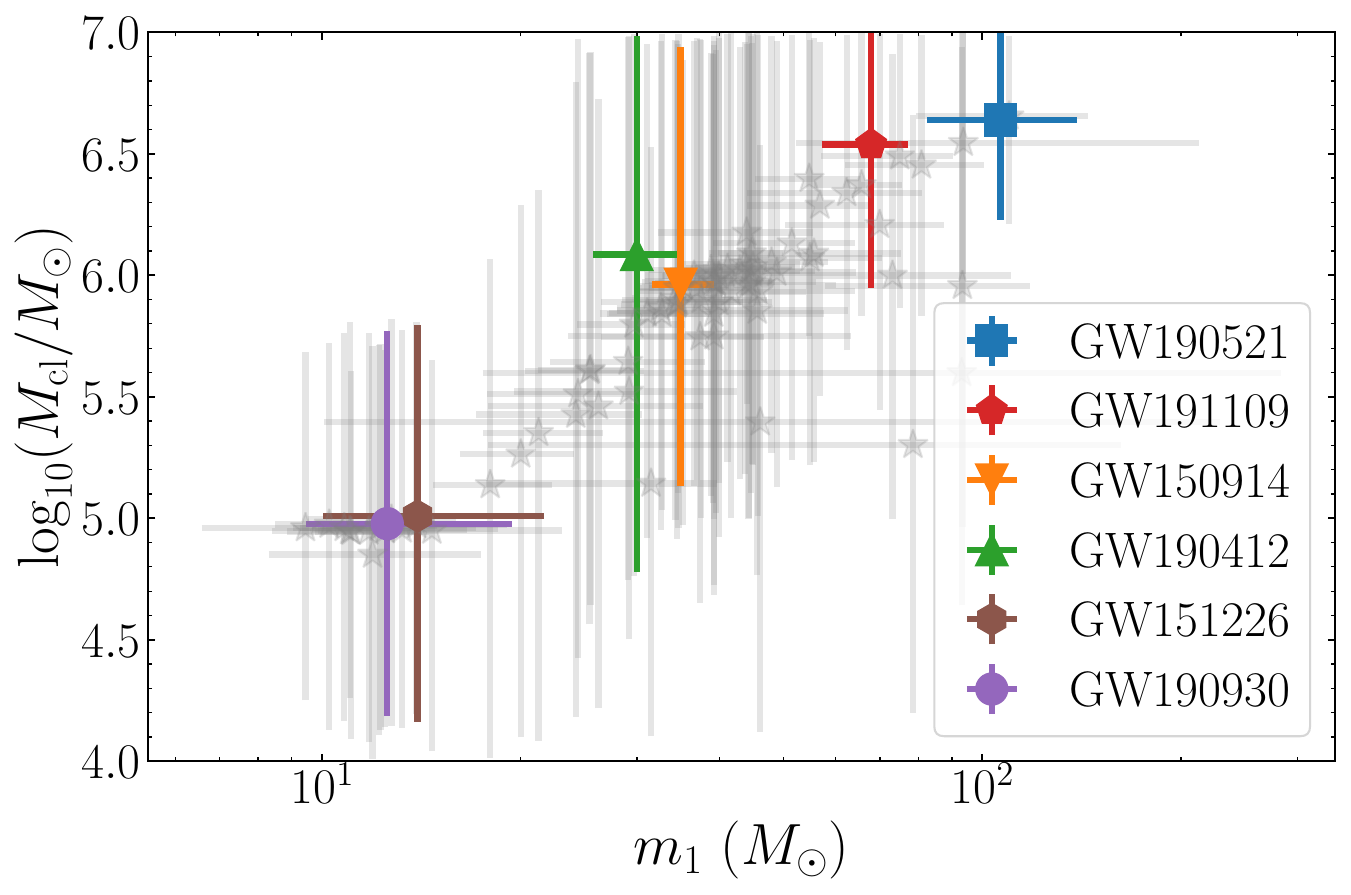}
    \caption{The 90\% credible intervals of the likelihood of $m_1$ (horizontal bars) and $\Mcl$ (vertical bars) for all BBH events in GWTC-3, with markers indicating the median values of $m_1$ and $\Mcl$.
    Six events (GW190521, GW191109, GW190412, GW150914, GW151226, and GW190930) are highlighted in color to better illustrate the correlation between $m_1$ and $\Mcl$.}
    \label{fig:selected}
\end{figure}

In Fig.~\ref{fig:corner} we show the joint likelihoods of $(\Mcl,\rh)$ for two events: GW190521~\cite{GW190521PE}, and GW151226~\cite{GW151226PE}.
These events were chosen because they have very different primary masses: GW190521 has $m_1\sim 100 \Msun$, suggestive of a hierarchical merger origin~\cite{GW190521astro}, while GW151226 has $m_1\sim 14 \Msun$, a more typical value for events in the GWTC-3 catalog~\cite{GWTC3rate}.

We find that the likelihood for $\rh$ is almost uninformative even for GW190521.
This is because, in the \textsc{Rapster} simulations~\cite{Kritos:2022ggc}, the compactness of the cluster mostly affects the number of BBHs produced in the cluster, i.e. $\etabbh$, rather than the shape of the BBH mass distribution.
As $\etabbh$ is not involved in the single-event cluster likelihood, we do not extract any new information about $\rh$ from the single-event analysis.
However, we note that the current version of \textsc{Rapster} does not include stellar mergers, which would allow for the possibility to form initial BHs within the PISN mass gap.
This mechanism may skew the distribution of $m_1$ to higher values for very compact clusters~\cite{Kremer:2020wtp}.

The key feature of Fig.~\ref{fig:corner} is that the likelihood of $\Mcl$ in the two systems is very different: GW190521 favors $\Mcl\gtrsim 10^6 \Msun$, and GW151226 favors $\Mcl\lesssim 10^6 \Msun$.
This may hint at a positive correlation between the primary BH mass $m_1$ and the probable cluster mass $\Mcl$ that produces the corresponding BBH merger event.

To test this hypothesis, we have analyzed all BBH events in GWTC-3.
The results of this analysis are shown in Fig.~\ref{fig:selected}, where we show the inferred value of $\Mcl$ as a function of $m_1$ for the GWTC-3 catalog. Most events are in grey, but a selected subset (listed in the legend) is highlighted in color. The highlighted subset is chosen to cover three different ranges of $m_1$: values in the PISN mass gap (GW190521 and GW191109, with $m_1\gtrsim40~\Msun$), events with $40 \gtrsim m_1\gtrsim 20\Msun$ (GW150914 and GW190412), and low-primary mass events with $m_1 \lesssim 20\Msun$ (GW151226 and GW190930).
As anticipated, we observe that the inferred values of $\Mcl$ (within the 90\% credible interval) tend to increase as a function of the measured values of $m_1$.
We have also checked that the correlation of $\Mcl$ with other binary parameters (such as $\chieff$ and $q$) is not as significant as the correlation with $m_1$.

\begin{figure}[t]
    \centering
    \includegraphics[width=0.9\columnwidth]{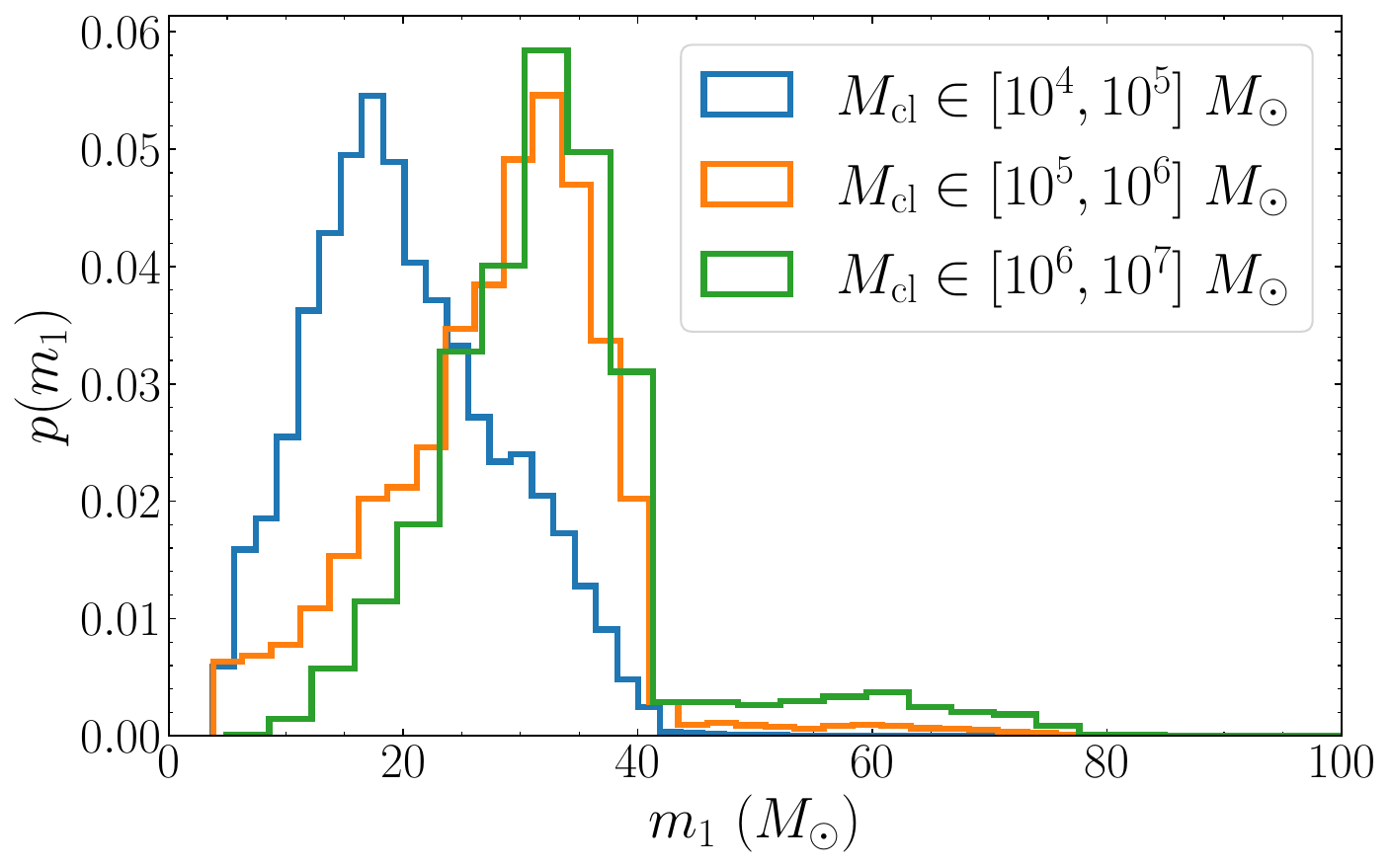}
    \caption{Histograms of the primary mass distributions generated by clusters with masses in the range $\Mcl \in [10^4, 10^5]~\Msun$ (blue), $[10^5, 10^6]~\Msun$ (orange), and $[10^6, 10^7]~\Msun$ (green), using a prior $p(\Mcl)\propto \Mcl^{-2}$.
    The peak location of the $m_1$ distribution, which represents the most probable first-generation mergers, shifts from $m_1\sim 15\Msun$ to $m_1\sim 35\Msun$ as $\Mcl$ increases from $10^4~\Msun$ to $10^6~\Msun$.}
    \label{fig:p_m1_compare}
\end{figure}

To understand this correlation, in Fig.~\ref{fig:p_m1_compare} we compare the primary mass distributions $p(m_1)$ generated by clusters having masses $\Mcl$ in different ranges (highlighted by histograms in different colors).
These mass distribution histograms have two major features.

First of all, the relative fraction of BBHs above the PISN mass gap is larger when $\Mcl \geq 10^5 \Msun$ (i.e., for the orange and green histograms). 
Hierarchical mergers within the mass gap occur more frequently in more massive clusters, because these clusters have larger escape velocities and thus they are more likely to retain the merger remnants despite their gravitational recoils.
This is compatible with the correlation between $\Mcl$ and primary BHs having $m_1\gtrsim 40\Msun$ observed in Fig.~\ref{fig:selected}.

Secondly, the mass distribution of first-generation mergers below the PISN mass gap (those with $m_1 \lesssim 40 \Msun$) is skewed toward lower values for $\Mcl \leq 10^5\Msun$ (blue histogram): for these light clusters, the peak in $m_1$ decreases from $\sim 35\Msun$ to $\sim 15\Msun$.
This trend may be qualitatively explained by a combination of the ejection mechanism discussed above, and the finite merging time window.
In a star cluster, first-generation mergers are typically formed by a combination of mass segregation and exchange interactions.
The majority of first-generation mergers are nearly equal-mass systems, whose critical semimajor axis for getting ejected out of the cluster after a binary-single interaction scales with $\aEj\propto m_1/\Mcl$: see e.g. Eq.~(8) in Ref.~\cite{Kritos:2022ggc}, or Eq.~(8) in Ref.~\cite{Antonini:2016gqe}.
Therefore, in less massive clusters the more massive BBHs are ejected at an earlier stage of their inspiral evolution.
Since the GW inspiral timescale $\tau\propto \aEj^4/m_1^3$~\cite{Peters:1964zz}, the typical inspiral time for ejected mergers has the scaling $\tau \propto m_1/\Mcl^4$.
As the binaries can only merge within the  (finite) cosmic time since their formation,
the critical $m_1$ below which BBHs can merge scales with $m_1 \propto \Mcl^4$. This leads to the observed shift in the primary mass distribution as $\Mcl$ decreases.
Note that this is only a qualitative explanation, and the quantitative correlation between $\Mcl$ and $m_1$ for first-generation BHs is very likely model-dependent.
The ejection efficiency and the resulting merging timescale are sensitive to nonlinear effects in cluster dynamics, to the formation redshift and to the cluster metallicity. All of these effects may modify the shape of the primary mass distribution at different merger redshifts.

\section{Discussion}
\label{sec:discussion}

In this paper we propose a two-level hierarchical framework to analyze the BBH population observed in GWs under the assumption that the merger events are produced dynamically in star clusters.
The two-level hierarchy is based on the idea that each cluster in the cluster population has different physical properties, and therefore produces a different BBH population.
The implication is that we can not only characterize the BBH population produced by an assumed cluster population, but (vice versa) we can also infer the population properties of the clusters, given a physical model of cluster dynamics.

To carry out this hierarchical inference, we first need to perform single-event inference -- that is, we need to identify the cluster properties that are most likely to produce any observed BBH merger event.
Estimating the single-event likelihood of cluster properties requires a knowledge of how the distribution of BBH parameters (such as binary masses, spins, and redshift) depends on the individual cluster properties, such as the cluster mass and radius. In this paper, we carried out a proof-of-principle demonstration that this hierarchical inference is possible using, for illustrative purposes, astrophysical models built with the \textsc{Rapster} code.
With \textsc{Rapster} we can simulate the formation of BBHs in star clusters, and then approximate the joint distribution of BBH parameters and cluster parameters from the simulated BBH samples by KDE methods.

We find that the inferred cluster mass is correlated with the measured BBH primary mass, as shown in Fig.~\ref{fig:selected}.
The correlation is a result of the variation of the primary mass distribution as a function of cluster mass observed in Fig.~\ref{fig:p_m1_compare}: more massive clusters enhance the production of hierarchical mergers above the PISN mass gap, and less massive clusters eject more massive first-generation binaries at semimajor axes that are too large to efficiently produce mergers.
As emphasized in the main text, this is a mostly qualitative explanation: the extent of the shift of the primary mass distribution in first-generation BBHs is sensitive to the details of cluster dynamics and to the initial conditions of cluster formation (including redshift and metallicity).

For cluster radii $\rh\in [0.5, 3]$~pc, the radius only affects the overall number of BBHs produced in each cluster (because stellar mergers are not been included in the model), and it has a negligible effect on the primary mass distribution.  In a more realistic scenario, the radius should play an important role in the inference of cluster population properties, because the number of BBHs produced per cluster $\etabbh$ entering the next hierarchy depends on $\rh$, and affects the expected BBH merger rate.
Our results are again model-dependent, and therefore it would be useful to validate this trend by comparing against other existing codes, especially for low-mass clusters with $\Mcl\lesssim 10^5\Msun$, whose evolution is more sensitive to the details of every dynamical process~\cite{Askar:2016jwt,Kritos:2022ggc,Joshi:1999vf,Rodriguez:2016kxx,Sedda:2021vjh}.
Another source of uncertainty is the initial binary fraction, which we assumed to be 10\% in the simulations. While differences in the initial binary star population have an impact on the exchange channel in \textsc{Rapster}, the effect of this population on the black hole mass-cluster mass correlation is probably more dependent on the physics of binary star evolution. The study of this effect would require complementing \textsc{Rapster} with input from a binary star evolution code to simulate common envelope and stable mass transfer for the original binary stars. This is an interesting problem, but it is beyond the scope of our work.

While the KDE method we employed suffices to capture the broad features discussed above, it may not be robust enough to proceed to the next hierarchy and infer the properties of the cluster population.
Performing the full hierarchical analysis may require more advanced techniques, such as deep generative modeling, to better approximate the multidimensional probability density functions involved in Eq.~\eqref{eq:clusterLikelihood}.
For example, one may learn $p(\xbbh|\xcl)$ and $\etabbh(\xcl)$ separately by simulating the BBH populations given a set of cluster properties $\{\xcl\}$, or work with the joint distribution $p(\xbbh, \xcl|\XclChosen)$ at a chosen cluster population characterized directly by $\XclChosen$, and obtain $p(\xbbh|\xcl)\etabbh(\xcl) \propto p(\xbbh, \xcl|\XclChosen)/p(\xcl | \XclChosen)$ by reweighing the chosen prior of the cluster population.

Finally, we note that the two-level hierarchy may be generalized to include contributions from multiple formation channels.
This may be achieved by using the relevant parametrization to set up Eq.~\eqref{eq:Nexpected} for each channel and build a mixture model in Eq.~\eqref{eq:fullHBA}.
For example, one may obtain the likelihood of progenitor redshift and metallicity based on binary evolution simulations for galactic field binaries (either by backpropagation or though other numerical techniques, as in Refs.~\cite{Wong:2022flg,Andrews:2017ads}); combine with the similar likelihood of cluster binaries; and trace the evolution of star formation rate, cluster formation rate, and stellar metallicity all at once.
The full hierarchical inference will be presented in future work.

\acknowledgments

We thank Davide Gerosa, Will Farr, Chase Kimball, Sharan Banagiri, and Vicky Kalogera for fruitful discussions.
KKYN is supported by Miller Fellowship at Johns Hopkins University and Croucher Fellowship by the Croucher Foundation.
K.K., A.A., R.C. and E.B. are supported by NSF Grants No. AST-2006538, PHY-2207502, PHY-090003 and PHY-20043, and NASA Grants No. 20-LPS20- 0011 and 21-ATP21-0010. This research project was conducted using computational resources at the Maryland Advanced Research Computing Center (MARCC).
This research has made use of data or software obtained from the Gravitational Wave Open Science Center (gwosc.org), a service of LIGO Laboratory, the LIGO Scientific Collaboration, the Virgo Collaboration, and KAGRA. LIGO Laboratory and Advanced LIGO are funded by the United States National Science Foundation (NSF) as well as the Science and Technology Facilities Council (STFC) of the United Kingdom, the Max-Planck-Society (MPS), and the State of Niedersachsen/Germany for support of the construction of Advanced LIGO and construction and operation of the GEO600 detector. Additional support for Advanced LIGO was provided by the Australian Research Council. Virgo is funded, through the European Gravitational Observatory (EGO), by the French Centre National de Recherche Scientifique (CNRS), the Italian Istituto Nazionale di Fisica Nucleare (INFN) and the Dutch Nikhef, with contributions by institutions from Belgium, Germany, Greece, Hungary, Ireland, Japan, Monaco, Poland, Portugal, Spain. KAGRA is supported by Ministry of Education, Culture, Sports, Science and Technology (MEXT), Japan Society for the Promotion of Science (JSPS) in Japan; National Research Foundation (NRF) and Ministry of Science and ICT (MSIT) in Korea; Academia Sinica (AS) and National Science and Technology Council (NSTC) in Taiwan.

\appendix
\section{Construction of KDE and importance sampling}
\label{App:KDEandSampling}

\textsc{Rapster} enables rapid simulations of BBHs generated from a population of clusters.
Therefore, it is more convenient to (i) choose a cluster population (parametrized by $\Tilde{\Lambda}_{\rm cl}$) that produces enough simulated BBHs in the range of the BBH parameter space  $\xbbh$ (our simulated samples), (ii) perform kernel density estimation (KDE) to approximate the joint distribution $\Tilde{p}(\xbbh, \xcl|\XclChosen)$ from the simulated mergers, and (iii) obtain $p(\xbbh|\xcl) = \Tilde{p}(\xbbh, \xcl|\XclChosen)/\Tilde{p}(\xcl|\XclChosen)$ from Bayes' theorem.

Here, $\Tilde{p}(\xcl|\XclChosen)$ contains a factor of the differential merger rate, $\etabbh(\xcl)$, on top of the chosen prior of the cluster population put into the simulations, since each input cluster may produce a different number of mergers.
To ensure that $p(\xbbh | \xcl)$ is properly normalized to unity for each $\xcl$, we require a second KDE for $\Tilde{p}(\xcl|\XclChosen) \propto \etabbh(\xcl) p(\xcl|\XclChosen)$, which can also be constructed from the simulated samples, because the count of $\xcl$ is proportional to the differential merger rate.
We employ \texttt{gaussian\_kde} from \texttt{scipy}, written in the \texttt{jax} infrastructure, to speed up the KDE. We use $\sim 7\times 10^5$ simulation points, with a bandwidth of $\sim 0.25$ set by Silverman's rule.

One may attempt to first approximate the integral of Eq.~\eqref{eq:clusterLikelihood} by an importance sum over the $\xbbh$ samples of individual BBH likelihoods, and then draw the $\xcl$ sample from the approximated $\xcl$ likelihood by Monte Carlo methods.
Since the parameter space of $\xcl$ has a relatively low dimension, we simplify the sampling procedures further by importance sampling of the joint distribution $p(d_i, \xbbh_i | \xcl_i) \equiv p(d_i | \xbbh_i) p(\xbbh_i | \xcl_i)$.
In practice, we append an additional set of $\xcl_i$ samples, $\{\xcl_{i,j}\}_{j=1}^K$, drawn from a uniform distribution $U(\xcl_i)\propto 1$ to the set of BBH likelihood samples, $\{\xbbh_{i,j}\}_{j=1}^K$.
The aggregated set $\{\xbbh_{i,j}, \xcl_{i,j}\}_{j=1}^K$ follows the joint distribution $p(d_i | \xbbh_i) U(\xcl_i)$.
The desired set of $\xcl_i$ samples that are representative of the marginalized likelihood $p(d_i | \xcl_i)$ is equivalent to the set of $\{\xcl_{i,j}\}_{j=1}^K$ weighed by $\{w^{\rm cl}_{i,j} \propto p(\xbbh_{i,j} | \xcl_{i,j})\}_{j=1}^K$.

\bibliography{references}

\end{document}